\documentclass[remotesensing,article,accept,moreauthors,pdftex]{mdpi} 

\firstpage{1} 
\pubvolume{12}
\issuenum{10}
\articlenumber{1685}
\pubyear{2020}
\copyrightyear{2020}

\history{Received: 28 March 2020; Accepted: 15 April 2020 ; Published: Waiting for conversions and typesetting}

\usepackage{amsmath,amssymb,amsfonts}
\usepackage{algorithmic}
\usepackage{graphicx}
\usepackage{multirow}
\usepackage{subfloat}
\usepackage{dblfloatfix}
\usepackage{epstopdf}
\usepackage{multirow}

\usepackage[caption=false]{subfig}

\Title{Classification of Arrhythmia by Using Deep Learning with 2-D ECG Spectral Image Representation}

\Author{{Amin Ullah}  $^{1,2}$, Syed Muhammad Anwar  $^{1,2}$ , Muhammad Bilal  $^{3}$ and Raja Majid Mehmood $^{4 *}$}

\AuthorNames{Amin Ullah, Syed Muhammad Anwar, Muhammad Bilal and Raja Majid Mehmmod}

\address{%
$^{1}$ \quad Software Engineering Department, University of Engineering and Technology Taxila, Punjab 47050, Pakistan; s.anwar@uettaxila.edu.pk  \\

$^{2}$ \quad Center for research in computer vision lab (CRCV Lab), College of Engineering and Computer Science, University of Central Florida (UCF), Orlando, FL 32816,  USA\\

$^{3}$ \quad Division of Computer and Electronics Systems Engineering,
Hankuk University of Foreign Studies, Yongin-si~17035,~Korea; m.bilal@ieee.org\\

$^{4}$ \quad Information and Communication Technology Department, School of Electrical and Computer Engineering, Xiamen University Malaysia, Sepang 43900, Malaysia. \\

}

\corres{Correspondence: rmeex07@ieee.org, rajamajid@xmu.edu.my}

\abstract{The electrocardiogram (ECG) is one of the most extensively employed signals used in the diagnosis and prediction of cardiovascular diseases (CVDs). The ECG signals can capture the heart's rhythmic irregularities, commonly known as arrhythmias. A careful study of ECG signals is crucial for precise diagnoses of patients' acute and chronic heart conditions. In this study, we propose a two-dimensional (2-D) convolutional neural network (CNN) model for the classification of ECG signals into eight classes; namely, normal beat, premature ventricular contraction beat, paced beat, right bundle branch block beat, left bundle branch block beat, atrial premature contraction beat, ventricular flutter wave beat, and ventricular escape beat. The one-dimensional ECG time series signals are transformed into 2-D spectrograms through short-time Fourier transform. The 2-D CNN model consisting of four convolutional layers and four pooling layers is designed for extracting robust features from the input spectrograms. Our proposed methodology is evaluated on a publicly available MIT-BIH arrhythmia dataset. We achieved a state-of-the-art average classification accuracy of 99.11\%, which is better than those of recently reported results in classifying similar types of arrhythmias. The performance is significant in other indices as well, including sensitivity and specificity, which indicates the success of the proposed method.}


\begin{document}

\setcounter{section}{0} 

\section{Introduction\label{sec1}}
Cardiovascular diseases (CVDs) are the leading cause of human death, with over 17 million people known to lose their lives annually due to CVDs \cite{mc2019cardiovascular}. According to the World Heart Federation, three-fourths of the total CVD deaths are among the middle and low-income segments of the society~\cite{1}. A classification model to identify CVDs at their early stage could effectively reduce the mortality rate by providing a timely treatment \cite{mustaqeem2020modular}. One of the common sources of CVDs is cardiac arrhythmia, where heartbeats are known to deviate from their regular beating pattern. A normal heartbeat varies with age, body size, activity, and emotions. In cases where the heartbeat feels too fast or slow, the condition is known as palpitations. An arrhythmia does not necessarily mean that the heart is beating too fast or slow, it indicates that the heart is following an irregular beating pattern. It could mean that the heart is beating too fast---tachycardia (more than 100 beats per minute (bpm)), or slow---bradycardia (less than 60 bpm), skipping a beat, or in extreme cases, cardiac arrest. Some other common types of abnormal heart rhythms include atrial fibrillation, atrial flutter, and ventricular fibrillation. 
These deviations could be classified into various subclasses and represent different types of cardiac arrhythmia. An~accurate classification of these types could help in diagnosing and treatment of heart disease patients. Arrhythmia could either mean a slow or fast beating of heart, or patterns that are not attributed to a normal heartbeat. An automated detection of such patterns is of great significance in clinical practice. There are certain known characteristics of cardiac arrhythmia, where the detection requires expert clinical knowledge.

The electrocardiogram (ECG) recordings are widely used for diagnosing and predicting cardiac arrhythmia for diagnosing heart diseases. Towards this end, clinical experts might need to look at ECG recordings over a longer period of time for detecting cardiac arrhythmia. The ECG is a one-dimensional (1-D) signal representing a time series, which can be analyzed using machine learning techniques for automated detection of certain abnormalities. Recently, deep learning techniques have been developed, which provide significant performance in radiological image analysis \cite{irmakci2020deep,anwar2018medical}. Convolutional neural networks (CNNs) have  recently been  shown to work for multi-dimensional (1-D, 2-D, and in certain cases, 3-D) inputs but were initially developed for problems dealing with images represented as two-dimensional inputs \cite{gu2018recent}. For time series data, 1-D CNNs are proposed but are less versatile when compared to 2-D CNNs. Hence, representing the time series data in a 2-D format could benefit certain machine learning tasks \cite{wu2018comparison,zhao2019speech}.  
Hence, for ECG signals, a 2-D transformation has to be applied to make the time series suitable for deep learning methods that require 2-D images as input. The short-time Fourier transform (STFT) can convert a 1-D signal into a 2-D spectrogram and encapsulate the time and frequency information within a single matrix. The 2-D spectrogram is similar to hyper-spectral and multi-spectral images (MSI), which have diverse applications in remote sensing and clinical diagnosis, including spectral un-mixing, ground cover classification and matching, mineral exploration, medical image classification, change detection, synthetic material identification, target detection, activity recognition, and surveillance \cite{1a,1b,1c,1d,1e,1f,1g}. The 2-D matrix of spectrogram coefficients could be useful for extracting robust features for representation of a cardiac ECG signal \cite{salem2018ecg}. This representation could allow the application of CNN architectures (designed to operate on 2-D inputs) for development of automated systems related to CVDs.   

\subsection{{Related Works}}

The ECG signal detects abnormal conditions and malfunctions by recording the potential bio-electric variation of the human heart. Accurately detecting the clinical condition presented by an ECG signal is a challenging task \cite{mustaqeem2017statistical}. Therefore, cardiologists need to accurately predict and identify the right kind of abnormal heartbeat ECG wave before recommending a particular treatment. This might require observing and analyzing ECG recordings that might continue for hours (patients in critical care). To overcome this challenge for the visual and physical explanation of the ECG signal, computer-aided diagnostic systems have been developed to automatically identify such signals automatically \cite{anwar2018arrhythmia}. Most of the research in this field has been conducted by incorporating different approaches of machine learning (ML) techniques for the efficient identification and accurate examination of ECG  signals \cite{mustaqeem2018multiclass,mustaqeem2017wrapper}. The ECG signal classification based on different approaches has been presented in the literature including frequency analysis \cite{2}, artificial neural networks (ANNs) \cite{3}, heuristic-based methods \cite{4}, statistical methods \cite{5}, support vector machines (SVMs) \cite{mustaqeem2018multiclass}, wavelet transform \cite{7}, filter banks \cite{8}, hidden Markov models \cite{9}, and mixture-of-expert methods \cite{10}. 
{An artificial neural network based method obtained an average accuracy of 90.6\% for the classification of ECG wave into six classes \cite{15}. Meanwhile, a feed-forward neural network was used as a classifier for the detection of four types of arrhythmia classes and achieved an average accuracy of 96.95\% \cite{16}.} 

Machine learning is a subset of artificial intelligence used with high-end diagnostic tools \cite{17,21,21a,21b} for the prediction and diagnosis of different types of illnesses \cite{22}. Deep learning, as a subset of ML, has many applications in the prediction and prevention of fatal sicknesses, particularly CVDs. Different techniques of deep learning used for the analysis of bioinformatics signals have been presented in \cite{13,17,19}. A recurrent neural network (RNN) \cite{20} was used for feature extraction and achieved an average accuracy of 98.06\% for detecting four types of arrhythmia. For the classification and extraction of features from a 1-D ECG signal, a 1-D convolutional neural network model was proposed \cite{24} and yielded a classification accuracy of 96.72\%. Another deeper 1-D CNN model was proposed for the classification of the ECG dataset \cite{25} and obtained an average accuracy of 97.03\%. In both instances, a large ECG dataset was used, but the ECG signals were represented as a 1-D time series. 
A nine-layer 2-D CNN model was applied for an automatic classification of five different heartbeat arrhythmia types achieving an accuracy of 94.03\% \cite{28}. 

\subsection{Our Contributions}
The conventional techniques might not achieve efficient results due to the inter-patient variability in ECG signals \cite{chen2019smart}. Additionally, the efficiency and accuracy of traditional methods could be negatively affected by the increasing size of data \cite{11,12,13}. The techniques presented in literature have been applied to smaller datasets; however, for the purpose of generalization, the performance should be tested on larger datasets. {There are methods reported that use 2-D ECG signals \cite{salem2018ecg,xiong2017robust}; however, to the best of our knowledge, there are not clear details on how the 1-D ECG signal is converted to 2-D images for using 2-D CNN models.} Most methods have been tested on only a few types of arrhythmia and must be evaluated on all major types of arrhythmia. It should be noted that the performance of methods developed for 1-D ECG signals can be further improved. Towards this end, the major contributions of our proposed work are:

\begin{enumerate}[leftmargin=*,labelsep=5mm]
	\item {Spectrograms (2-D images) are employed, which are generated from the 1-D ECG signal using STFT. In addition, data augmentation was used for the 2-D image representation of ECG signals.}
	\item {A state-of-the-art performance was achieved in ECG arrhythmia classification by using the proposed CNN-based method with 2-D spectrograms as input.}

\end{enumerate}

The rest of the paper is organized as follows. The proposed algorithm is presented in detail in Section \ref{sec2}. The experiments conducted for the validation of the proposed scheme is presented in Section~\ref{sec3}. Classification results are presented in Section \ref{sec4}, and conclusions in Section \ref{sec5}.


\section{Proposed Scheme\label{sec2}}

\begin{figure}[H]
	\centering
	\includegraphics[width=130mm]{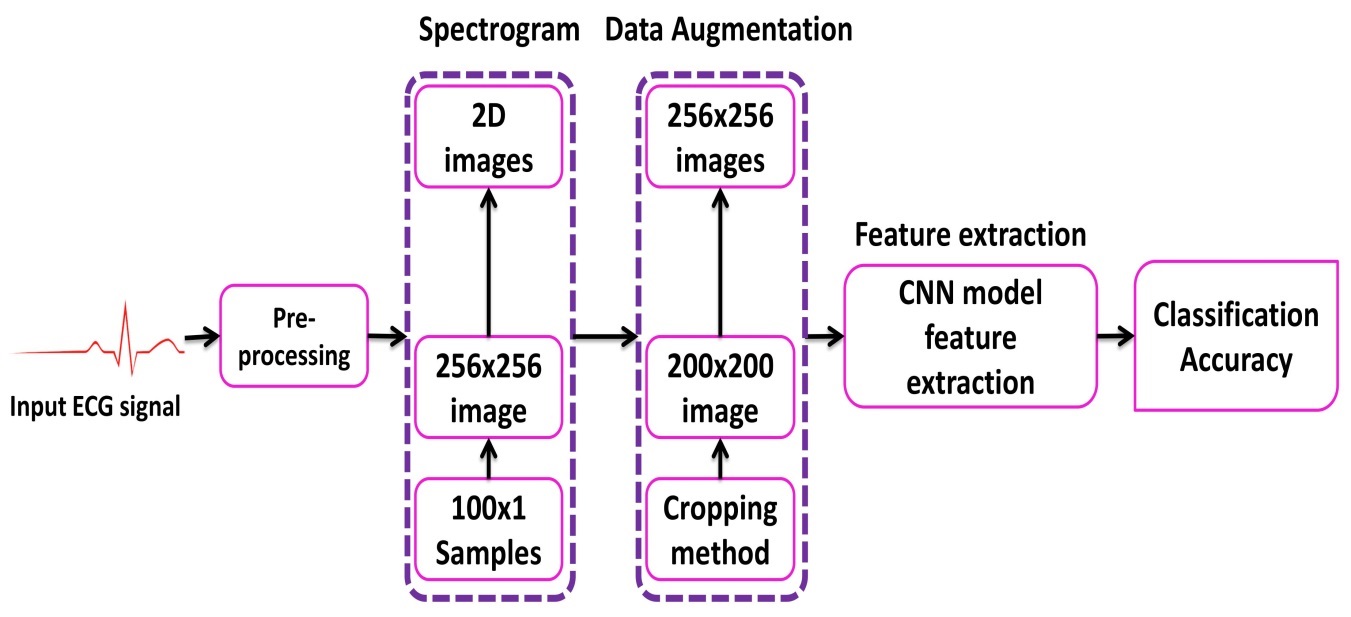}
	\caption{Complete procedure of electrocardiogram (ECG) signal classification.\label{fig1}}
\end{figure}

{A schematic representation of the proposed scheme is presented in Figure \ref{fig1}. The method consists of five steps, i.e., signal pre-processing, generation of 2-D images (spectrograms), augmentation of data, extraction of features from the data (using the CNN model), and its classification based on the extracted features. The details of these steps are presented in the following subsections.} 

\vspace{0.05cm}

\subsection{Pre-Processing}

The three primary forms of noise in the ECG signal are power line interference, baseline drift, and electromyographic noise \cite{34}. The noise from the original ECG signal must be removed to ensure that a denoised ECG signal is obtained for further processing. {We combined wavelet based thresholding and the reconstruction algorithm of wavelet decomposition to remove noise from the original ECG signal \cite{35}. The wavelet thresholding was performed using,} 

\begin{equation}
    {\overline{ \omega_{x, y}}}=\left\{\begin{array}{ll}\operatorname{sgn}\left(\omega_{x, y}\right)\left\{\left|\omega_{x, y}\right|-\frac{\lambda}{\exp ^{3}\left[\alpha\left(\left|\omega_{x, y}\right|-\lambda\right) / \lambda\right]}\right\}, & \left|\omega_{x, y}\right| \geq \lambda \\ 0, & \left|\omega_{x, y}\right|<\lambda\end{array}\right. 
\end{equation}
{where $w_{x,y}$ represents the wavelet coefficients, $\overline{ \omega_{x, y}}$ represents the estimated wavelet coefficients after threshold, $x$ represents the scale and $y$ represents the shift, $\lambda$ represents the threshold, and $\alpha$ is a parameter whose value can be set arbitrarily. The wavelet thresholding reduced the electromyographic noise and power line noise interference. Moreover, the reconstruction algorithm of wavelet decomposition was used to remove the baseline drift noise from the noisy ECG signal.} 

\subsection{Generation of 2-D Images}
{While 1-D CNN can be used for time series signals, the flexibility of such models is limited due to the use of 1-D kernels. On the other hand, 3-D CNNs require a large amount of training data and computational resources. In comparison, 2-D CNNs are more versatile since they use 2-D kernels and, hence, could provide representative features for time series data. Hence, for certain applications where sufficient data is available and for 1-D signals that can be represented in a 2-D format, using a 2-D CNN could be beneficial. Herein, for generating 2-D images to be used with the 2-D CNN model, the ECG signal was transformed into a 2-D representation.} 
{The 2-D time-frequency spectrograms were generated using the short-time Fourier transform. The ECG signal represents non-stationary data where the instantaneous frequency varies with time. Hence, such changes cannot be fully represented by just using information in the frequency domain. The STFT is a method derived from the discrete Fourier transform to analyze instantaneous frequency as well as the instantaneous amplitude of a localized wave with time-varying characteristics. In the analysis of a non-stationary signal, it is assumed that the signal is approximately stationary within the span of a temporal window of finite support. The 1-D ECG signals were converted into 2-D spectrogram images by applying STFT as~follows,}
\begin{equation}
    X_{STFT}[m, n]=\sum_{k=0}^{L-1} x[k] g[k-m] e^{-j 2 \pi n k / L} 
\end{equation}
{where $L$ is the window length, and $x[k]$ is the input ECG signal. The log values of $X_{STFT}[m,n]$ are represented as spectrogram (256 $\times$ 256) images.} 

\subsection{Data Augmentation}
{Another significant advantage of using 2-D CNN models is the flexibility it provides in terms of data augmentation. For 1-D ECG signals, data augmentation could change the meaning of the data and hence is not beneficial. However, with 2-D spectrograms, the CNN model can learn the data variations, and augmentation helps in increasing the amount of data available for training. The ECG data is highly imbalanced, where most of the instances represent the normal class. In this scenario, data augmentation can help when those classes that are underrepresented are augmented. 
For arrhythmia classification using ECG signals, augmenting training data manually could degrade the performance. Moreover, classification algorithms such as SVM, fast Fourier neural network, and tree-based algorithms, assume that the classification of a single image based representation of an ECG signal is always the same~\cite{26}. The proposed CNN model works on 2-D images of ECG signals as input data, which allows changing the image size with operations such as cropping. Such augmentation methods would add to the training data and hence would allow better training of the CNN model. 
Another important issue that arises when using small data with CNN based architectures is overfitting. Data augmentation is a way to deal with overfitting and allows better training of a CNN model. 
For imbalanced data, data augmentation can help in maintaining a balance between different classes. 
We have used the cropping method for the augmentation of seven classes of ECG beats; namely, premature ventricular contraction beat (PVC), paced beat (PAB), right bundle branch block beat (RBB), left bundle branch block beat (LBB), atrial premature contraction beat (APC), ventricular flutter wave (VFW), and ventricular escape beat (VEB). These are common types of cardiac arrhythmias and are considered in studies we have used for comparison (refer to the Discussion section). 
While other methods of augmentation are used, such as warping in image processing applications, the aim here is to augment classes that are under-represented. Towards this end, eight different cropping operations (left top, center top, right top, left center, center, right center, left bottom, center bottom, right bottom) were applied. As a result of cropping, we obtain multiple ECG spectrograms of reduced size (200 $\times$ 200), which are then resized to 256 $\times$ 256 images (using linear interpolation) before being fed into the CNN. This resulted in an eight times increase in the training data, which benefited the training process.}

\subsection{Deep Neural Network}

{In this study, a CNN-based model is proposed for an automatic classification of arrhythmia using the ECG signal in a supervised manner. The ECG data used in the study have corresponding labels (ground truth) identifying the type of arrhythmia present. These labels were assigned by expert cardiologists and are used for supervised training. For each heartbeat segment, the arrhythmia class label was transferred to the corresponding spectrogram image representation.} The first CNN-based algorithm, introduced in 1989 \cite{36}, was developed and used for the recognition of handwritten zip codes. {Since then, multiple CNN models have been proposed for the classification of images, among which AlexNet \cite{24} has achieved significant performance for a variety of images. The existing neural networks with the feed-forward process for the automatic classification of the 2-D image was not feasible since these methods do not take into account the local spatial information. 
However, with the development of CNN architectures and using nonlinear filters, spatially adjacent pixels can be correlated to extract local features from the 2-D image. In the 2-D convolution algorithm, the downsampling layer is highly desirable for extracting and filtering the spatial vicinity of the 2-D ECG images. For these reasons, the ECG signal was transformed into a 2-D representation, and a 2-D CNN algorithm was used for classification.} Consequently, high accuracy was obtained in the automatic taxonomy of the ECG waves. 
{The details of the proposed CNN model is presented in Section \ref{sec3.2}.}

\begin{figure}[H]
	\centering
	\includegraphics[width=142mm]{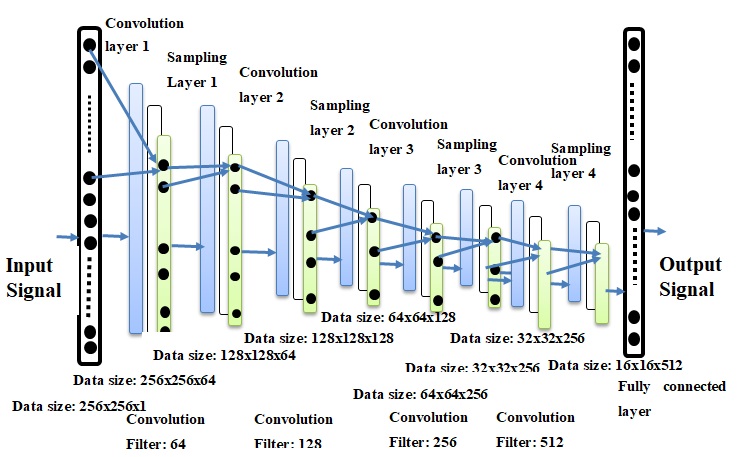}
	\caption{The architecture of the proposed convolutional neural network (CNN) model. architecture.\label{fig2}}
\end{figure}
\vspace{0.05cm}

\section{Experiments\label{sec3}}
\unskip
\subsection{Dataset\label{sec3.1}}
{The MIT-BIH arrhythmia dataset consists of 48 records, each having an approximate duration of 30 minutes recorded from a two-channel ambulatory system, collected between 1975 and 1979 \cite{moody2001impact}. Twenty-three recordings were selected at random from 4000 long term Holter recordings composed of a diverse group of inhabitants of indoor patients (60\%) as well as outdoor patients (40\%). Twenty-five recordings were chosen from a similar set, with a focus on complex ventricular, junctional, and supra-ventricular arrhythmias. These recordings were digitized at 360 samples/sec for each channel with a resolution of 11-bits over a range of 10 mV. A minimum of two cardiologists were involved in annotating each record and recorded the issues and corresponding solutions needed to reach to the computer-readable outcome. Hence, for the records, approximately 110,000 explanations were documented in this database. The data is publicly available for download here: \url{https://www.physionet.org/content/mitdb/1.0.0/}.}

\subsection{Deep Neural Network Parameters\label{sec3.2}}
The performance of the proposed CNN algorithm was compared with AlexNet and VGGNet architectures \cite{26} in terms of the ECG arrhythmia classification. 
{The regular normal beat (NOR) and seven other types of cardiac arrhythmia (VFW, PVC, VEB, RBB, LBB, PAB, and APC) classes were selected from the MIT-BIH arrhythmia database. Although the data is annotated with eighteen different classes, some of the classes have extremely low representation. Moreover, the selected eight types are more commonly found (hence having acceptable representation in the ground truth data) and also used by the methods we have evaluated for comparison. The architecture of the CNN model used in our experiments is shown in Figure \ref{fig2}. A detailed representation of layers within the model are presented in Table \ref{Table1}. The model follows the CNN architecture with four 2-D convolutional layers. Each convolutional layer is followed by a pooling layer. The output layer is a softmax layer with eight neurons to give the final classification. A fully connected layer is used between the last pooling layer and the output layer and represents the features learned by the CNN model.}   

\begin{table}[H]
	\caption{Details of the layers used in the proposed CNN model architecture.} 
	\centering 
	\renewcommand{\arraystretch}{1.2}
	\scalebox{0.95}{%
		\begin{tabular}{ccccccc} 
			\toprule
			\textbf{ Layers}  &	\textbf{Type} &	\textbf{Filter Size}	& \textbf{Stride}  &	 \textbf{Kernel} &	\textbf{Input Size} & \textbf{Parameters} \\ \midrule 
			
		Layer 1	& Conv2-D & 	3 $\times$ 3 & 	1 & 	64 & 	256 $\times$ 256 $\times$ 1 & 576\\
		Layer 2	& Pooling & 	2 $\times$ 2	 & 2 & 	- & 	256 $\times$ 256 $\times$ 64 & -\\
		Layer 3	& Conv2-D & 	3 $\times$ 3	 & 1 & 	128 & 	128 $\times$ 128 $\times$ 64 & 73,728\\
		Layer 4	& Pooling & 	2 $\times$ 2	 & 2 & -	 & 	128 $\times$ 128 $\times$ 128 &-\\
		Layer 5	& Conv2-D & 	3 $\times$ 3	 & 1 & 	256 & 	64 $\times$ 64 $\times$ 128 & 294,912\\
		Layer 6	& Pooling & 	2 $\times$ 2	 & 2 & -	 & 	64 $\times$ 64 $\times$ 256 & -\\
		Layer 7	& Conv2-D & 	3 $\times$ 3	 & 1 & 	512	 & 32 $\times$ 32 $\times$ 256 & 1,179,648 \\
		Layer 8	& Pooling  & 	2 $\times$ 2  & 	2 & -&		32 $\times$ 32 $\times$ 512 & -\\
		Layer 9	& Fully Connected & -	 & 	- & 	4096 & 	16 $\times$ 16 $\times$ 512 & 2,097,152\\
		Layer 10 & 	Output Layer & 	- &-  & 		8 	 & 4096 & 32,776\\	
			\bottomrule
		\end{tabular}}
		\label{Table1} 
	\end{table}

\subsection{Experimental Setup}
The proposed CNN classifier was implemented in Python with the open source library Tensor Flow \cite{38}, which was developed by Google for deep learning. Substantial computational power and training time were needed to train the CNN model. 
{The experimental setup consisted of an eighth-generation ASUS server with 32GB internal RAM, 500 GB external SSD hard drive with the addition of internal hard drive, and NVIDIA 1080 GPU with 11GB memory. The 2-D spectral images were divided such that 70\% of the data was used for training, 30\% for test. A 5-fold cross validation was used during the training process. The train/test splits were generated such that there was no overlap between the two splits.}   

\subsection{Cost Function}
The cost function is used to measure the error of the CNN model between the estimated worth and the actual worth or the desired quality. An optimizer function was used to minimize the error function. Different cost functions have been used in the neural network theory. In our experiments, we~used the cross-entropy function which is given as,

\begin{equation}
C = \frac{-1}{n} \sum_{c=1}^{N} ([ y_c * ln(a_c)+(1-y_c)  ln(1-a_c)])  \label{equation2}                          
\end{equation}
{where $C$ represents the cost that needs to be minimized, $n$ is the number of training points, $y$ is the expected or target value, N is the total number of classes, c is the class index, and $a$ is the actual value.} A gradient descent algorithm was used as an optimizer function with a learning rate of $0.001$ to reduce the error of cost function. 
Adam optimizer was used in the experiments for training the proposed CNN model, and it reached the optimal point in fewer iterations.  

\subsection{Evaluation Parameters }

{Four evaluation metrics were used in this study, including accuracy, precision, sensitivity, and specificity. The accuracy for the multi-class problem was evaluated as,}

\begin{equation}
A  = \frac{1}{N} \sum_{c=1}^{N} \frac{(T_p^c+T_N^c)}{(T_P^c+T_N^c+F_P^c+F_N^c )}, \label{equation3} 
\end{equation}
{where
$T_p$ denotes the true positives,
$F_p$ represents the false positives,
$T_N$ represents the true negatives, and
$F_N$ represents the false negatives, c represents the class index, and N represents the total number of classes. The accuracy (A) represents the ratio of the correctly classified instances to that of the total number of instances. The precision ($P$) and sensitivity ($Sen$) were calculated as,}

\begin{equation}
P  =  \frac{1}{N} \sum_{c=1}^{N} \frac{T_p^c}{T_p^c+ F_p^c},     \label{equation4}                           
\end{equation}

\begin{equation}
 Sen  =\frac{1}{N} \sum_{c=1}^{N}  \frac{T_p^c}{T_p^c+ F_N^c }.   \label{equation5}                                 
\end{equation}

{The specificity (Sp), also known as the true negative rate, was calculated as,} 
             
\begin{equation}
Sp       = \frac{1}{N} \sum_{c=1}^{N}  \frac{T_N^c}{T_N^c+F_P^c}.       \label{equation6}                                    
\end{equation}   

{The F1 score was calculated using the precision (P) and recall (Sen) as,}
\begin{equation}
F1 Score      = 2 \times (\frac{P \times Sen}{P + Sen}). 
\label{fscore}                                    
\end{equation}   

\section{Classification Results and Discussion}\label{sec4}
\unskip
\subsection{Results}

The two significant optimization parameters in the proposed 2-D CNN model are the learning rate and the batch size of the data used. To improve the performance, these two optimization parameters must be selected carefully to obtain the best accuracy in the automatic classification of arrhythmia using the ECG signals.
The proposed model was evaluated in different experiments with various values of learning parameters. For a smaller value of the learning rate (i.e., less than 0.0005), the speed of the convergence was very slow. However, when the value of the learning rate was large (i.e., greater than 0.001), the speed of convergence improved. At the same time, asymmetrical changes were observed in the accuracy rate. Henceforth, we selected an optimum value of 0.001 for the learning rate, as this value can attain better accuracy for the proposed model (i.e., optimum value), as shown in Table \ref{Tablebatch}.

\begin{table}[H]
	\caption{Batch sizes and average accuracy for a learning rate of 0.001.} 
	\centering 
	\renewcommand{\arraystretch}{1.23}
	\scalebox{0.9}{%
		\begin{tabular}{ccc} 
		\toprule
			\textbf{Learning Rate} &	\textbf{Batch Size}  &		\textbf{Average Accuracy}\\ \midrule
			0.001 &		2800 &		99.11\\
			0.001 &		2000 &		98.96\\
			0.001 &		1000 &		99.00\\
			0.001 &		500  &		98.95\\
			0.001 &		100  &		98.93\\
			
		\bottomrule
		\end{tabular}}
		\label{Tablebatch} 
	\end{table}

Similar to the learning rate, the batch size of the data also greatly affected the behavior and accuracy of the model. When the batch size was chosen to be 1000, the accuracy of the system showed abnormally large fluctuations in terms of system convergence. When the batch size was set to 2000, the accuracy of the system increased but did not reach a stable state. When the batch size was further increased to 2800, the accuracy of the proposed model was the highest and reached a stable state. The results are summarized in Table \ref{Tablelearn}. 

\begin{table}[H]
	\caption{Learning rate and average accuracy for a batch size of 2800.} 
	\centering 
	\renewcommand{\arraystretch}{1.23}
	\scalebox{0.9}{%
		\begin{tabular}{ccc } 
		\toprule
\textbf{Batch Size} & 	\textbf{Learning Rate}  & 	\textbf{Average Accuracy}\\ \midrule
2800  & 	0.001  & 	99.11\\
2800  & 	0.005  & 	98.84\\
2800  & 	0.100    & 	98.89\\
2800  &   	0.200	   &     98.91\\

		\bottomrule
		\end{tabular}}
		\label{Tablelearn} 
	\end{table}

\begin{figure}[H]
		\centering
		\includegraphics[width=105mm]{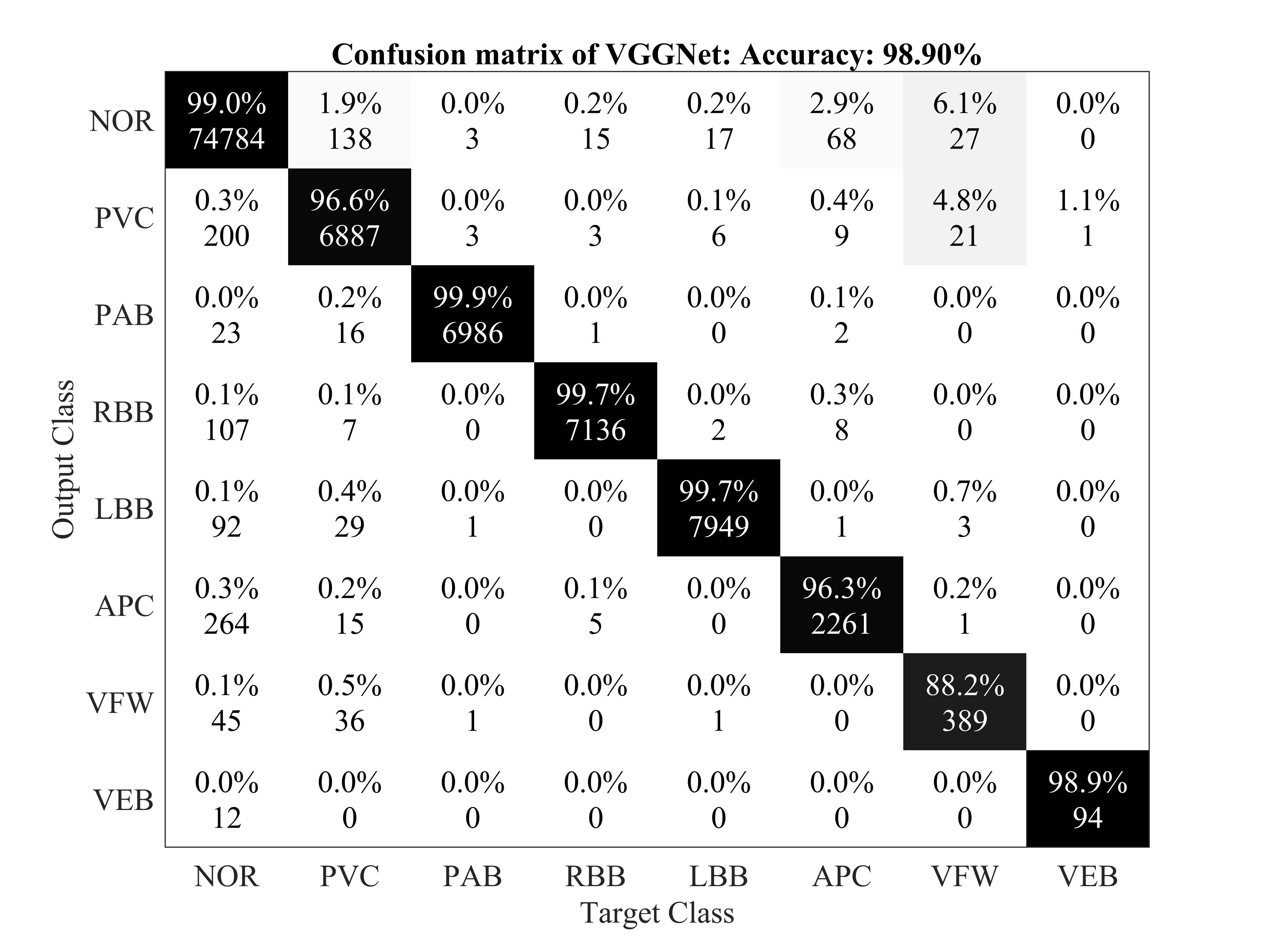}
		\caption{Confusion matrix for VGGNet.\label{fig3}}
	\end{figure}

\begin{figure}[H]
		\centering
		\includegraphics[width=105mm]{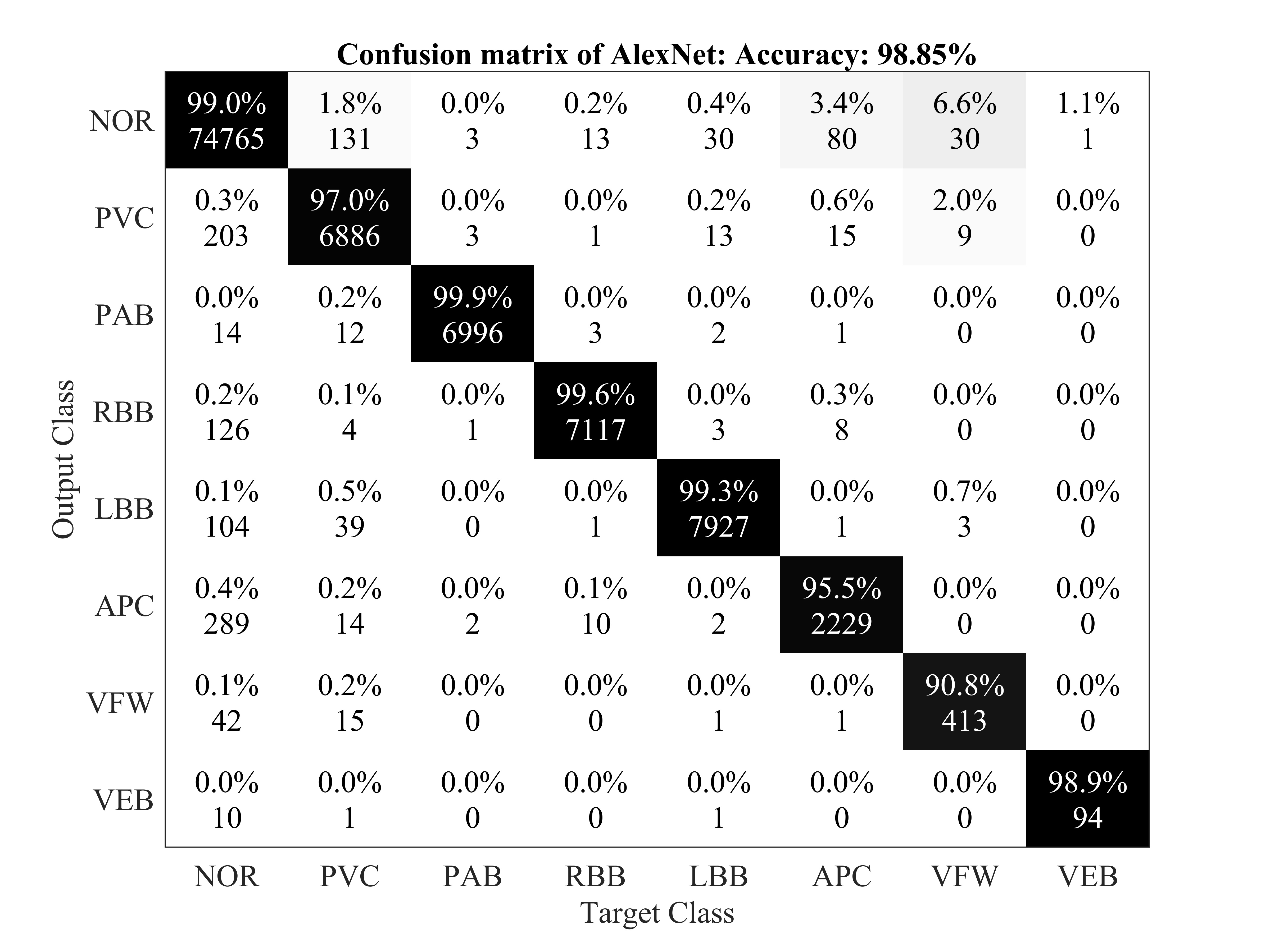}
		\caption{Confusion matrix for AlexNet.\label{fig4}}
	\end{figure}

\begin{figure}[H]
		\centering
		\includegraphics[width=105mm]{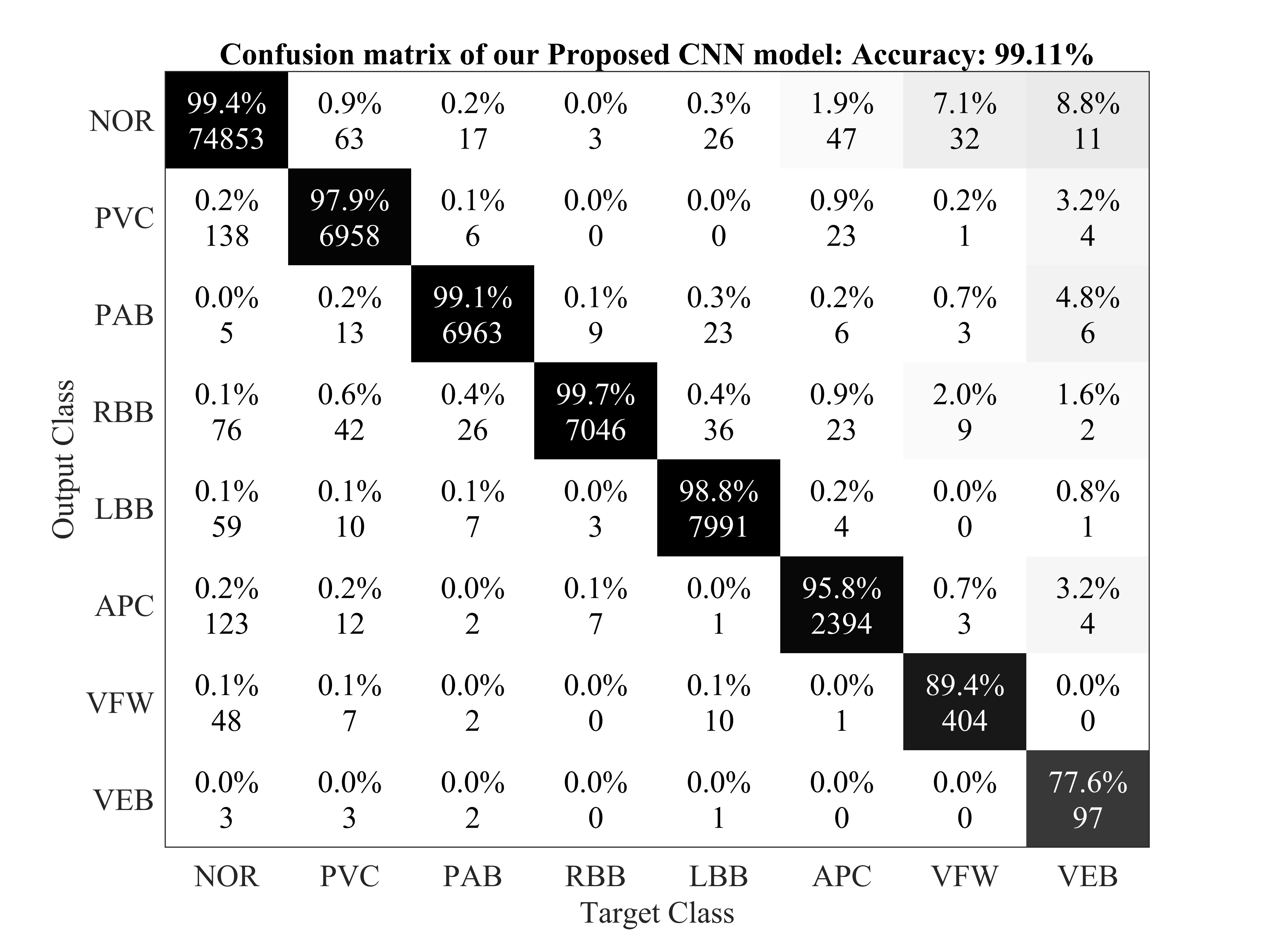}
		\caption{Confusion matrix for the proposed 2-D CNN based classification model.\label{fig5}}
	\end{figure}

{A detailed performance comparison between the proposed 2-D CNN model and other CNN models (including VGGNet and AlexNet) is presented using confusion matrices for all eight classes. The diagonal elements show the correctly classified classes, whereas anything off diagonal represents an incorrect classification. For the 2-D ECG data used in experiments, results are presented for the VGGNet (Figure \ref{fig3}), AlexNet (Figure \ref{fig4}), and the proposed model (Figure \ref{fig5}). The average accuracy of these three models is presented by averaging the diagonal values.} 

	\vspace{0.05cm}

	\vspace{0.05cm}

	\vspace{0.05cm}

\subsection{Discussion}
Table \ref{Table2} summarizes the performance evaluation of the proposed CNN algorithm with other classification methods of arrhythmia using ECG signals. The terms 'native' and 'augmented' in Table \ref{Table2} represent the training set without and with data augmentation, respectively. However, a direct comparison of our proposed CNN model with existing techniques may be unfit due to variations in the training and testing dataset, size of the ECG dataset used for experiments, architecture of the CNN models used, and the varying number of types of arrhythmia used for classification. {It should be noted that there are various methods that used 1-D data directly for the classification of arrhythmia \cite{42,43,44,45,46,48,49}. Among these methods, 1-D CNN models have been proposed with a lower classification accuracy (\cite{46}---96.40\% and \cite{49}---93.60\%)
~when compared with the proposed model. We also used 1-D ECG signals as input to the CNN model used in experiments and achieved a classification accuracy of 97.80\%. In recent years, 2-D CNN models have also been used, by converting the 1-D ECG signals to 2-D representation, with noticeable performance \cite{salem2018ecg}. Towards this end, the proposed model was based on a 2-D representation of the ECG data to efficiently apply 2-D CNN models and benefit from the flexibility of data augmentation in such methods.}    

The proposed 2-D CNN model attained better accuracy, sensitivity, and specificity (in eight class classification) than the FFNN \cite{39} model, which classified only four kinds of arrhythmia. 
{It~was observed that the VGGNet model performs worse than the proposed model, albeit a deeper network. One of the reasons for these observations could be the deeper architecture of VGGNet and limited training data.    
}These results prove that the proposed CNN model has the state-of-the-art accuracy for the automatic classification of arrhythmia based on the comparison with different CNN based algorithms. Varying performance among the compared CNN models is due to the difference in their architectures and the number of convolution filters used in these CNN models' structures. In the proposed CNN model, we employed four convolutional layers, four downsampling (pooling) layers, and one fully connected layer. In the AlexNet model, six convolutional layers, three downsampling layers, and two fully connected layers were used, while the VGGNet model entailed ten convolutional layers, four downsampling layers, and two fully connected layers. By adding a convolutional or a downsampling layer to the architecture of the CNN models, the computational resources and the simulation time for training and testing the models also increase, and this is the main reason for using a carefully selected CNN model. {Since we have a limited amount of data, more deeper networks (such as DenseNet or ResNet) would not qualify to perform well within the scope of this problem. The proposed model can be trained on other classes of arrhythmia, although we did not perform this analysis so that we can compare our work with published results that use a 2-D representation of ECG~data.}

\begin{table}[H]
	\caption{Comparison of the proposed model with state-of-the-art ECG classification techniques.} 
	\centering 
	\renewcommand{\arraystretch}{1.15}
	\scalebox{0.87}{%
		\scalebox{.95}[.95]{\begin{tabular}{lccccccc} 
			\toprule
			\textbf{Model} &	\textbf{Native/Augmentation} &	\textbf{Classes}	& \textbf{Accuracy \%}	& \textbf{Sensitivity \%} &	\textbf{Specificity \%} &	\textbf{Precision \% }& \textbf{F1 Score} \\ \midrule
			FFNN \cite{39}&		             &	 		4  &		96.94 &		96.31 &		97.78 	&	- & -\\  \midrule
			PNN \cite{40}	     	 &	 &	 		8  &		98.71 &		- &		99.65 &	-  & - \\ \midrule
			SVM \cite{41}	                 &	 &	 		6  &		91.67 &		93.83 &		90.49 &		- & - \\  \midrule
			RNN \cite{42}                 &	 &	 		4  &		98.06 &		98.15 &		97.78 &		- & - \\  \midrule
			LS-SVM \cite{43}	 &	 &	 		3  &		95.82 &		86.16 &		99.17 &		97.01 & 0.91 \\  \midrule
			RFT \cite{44} 	                     &	  	  &		3  &		92.16 &		-     &		-     &		- & -\\  \midrule
			KNN \cite{45}	                     &	  	  &		17 &		97.00	  &	    96.60  &		95.80  &		- & - \\  \midrule
			1-D CNN \cite{46}                                         &	 &	 		5  &		96.40  &		68.80  &		99.50  &		79.20 & 0.73 \\  \midrule
			AlexNet \cite{26}                                         &	Augmented &	 		8  &		98.85 &		97.08 &		99.62 &		98.59 & 0.97\\  \midrule
			AlexNet \cite{26}                                         &	Native &	 		8  &		98.81	 &		96.81	 &		99.68	 &		98.63 & 0.97 \\  \midrule
			VGGNet \cite{26}                                         &	Augmented &	 		8  &		98.63 &		96.93 &		99.37 &		97.86 & 0.97\\  \midrule
			
			VGGNet \cite{26} & Native	 & 8 &98.77	& 97.26	& 99.43	& 98.08 &0.97\\ \hline
			2-D CNN \cite{48}											 &	      &		5  &		97.42 &		-     &		-	  &	    - \\  \midrule
			1-D CNN \cite{49}											 &	      &		7  &		93.60  &    	-     &	-         &	- \\  \midrule
			\textbf{Proposed (1-D)}		                                     &	 Native    &			\textbf{8}  &		\textbf{97.80} &		- &		- &		- & -\\
			 \midrule
			\textbf{Proposed (2-D)}		                                     &	 Augmented    &			\textbf{8}  &		\textbf{99.11} &		\textbf{97.91} &		\textbf{99.61} &		\textbf{98.58} & \textbf{0.98}\\
			 \midrule
			\textbf{Proposed (2-D)}		                                     &	 Native    &			\textbf{8}  &		\textbf{98.92} &		\textbf{97.26} &		\textbf{99.67} &		\textbf{98.69} & \textbf{0.98}\\
			
			\bottomrule 
		\end{tabular}}}
		\label{Table2} 
	\end{table}
We compared the proposed CNN-based model with recent techniques for the automatic classification of arrhythmia (Table \ref{Table2}), where the algorithm achieved 97.88\% average sensitivity, 99.61\% specificity, 99.11\% average accuracy, and 98.59\% positive predictive value (precision). These values indicate improved performance when compared with recent methods using of 1-D and 2-D CNNs, given the same arrhythmia classification. The results also show that the proposed CNN algorithm has better results in terms of accuracy with both the augmented and without augmented data. The~proposed model has attained the highest sensitivity among all the compared CNN algorithms. 
{It is pertinent to note that detecting these cardiac arrhythmias is a labor intensive task, where a clinical expert needs to carefully observe recordings that can go for up to hours. With such automated methods, the artificially intelligent system could augment the performance of clinical experts by detecting these patterns and directing the observer to look more closely at regions of more significance. This would ultimately improve the clinical diagnosis and treatment of some of the major CVDs.}

\section{Conclusions\label{sec5}}
In this study, we proposed a 2-D CNN-based classification model for automatic classification of cardiac arrhythmias using ECG signals. An accurate taxonomy of ECG signals is extremely helpful in the prevention and diagnosis of CVDs. Deep CNN has proven useful in enhancing the accuracy of diagnosis algorithms in the fusion of medicine and modern machine learning technologies. The~proposed CNN-based classification algorithm, using 2-D images, can classify eight kinds of arrhythmia, namely, NOR, VFW, PVC, VEB, RBB, LBB, PAB, and APC, and it achieved 97.91\% average sensitivity, 99.61\% specificity, 99.11\% average accuracy, and 98.59\% positive predictive value (precision). These results indicate that the prediction and classification of arrhythmia with 2-D ECG representation as spectrograms and the CNN model is a reliable operative technique in the diagnosis of CVDs. The~proposed scheme can help experts diagnose CVDs by referring to the automated classification of ECG signals. 
The present research uses only a single-lead ECG signal. The effect of multiple lead ECG data to further improve experimental cases will be studied in future work.

\vspace{6pt} 
\authorcontributions{Conceptualization, A.U and M.A; Methodology, A.U, R.M, M.A; Validation, A.U, R.M and M.B; Formal Analysis, A.U, M.A; Writing- Original Draft Preparation, A.U, R.M, M.B, M.A; Writing- Review \& Editing, M.A, A.U; Supervision, M.A; Funding Acquisition, R.M}


\funding{This research has funded by the Xiamen University Malaysia Research Fund (XMUMRF) (Grant No: XMUMRF/2019-C3/IECE/0007).}


\acknowledgments{The authors thank for the valuable advice from Prof. Ulas Bagci (Center for Research in Computer Vision (CRCV) Laboratory, University of Central Florida (UCF),  Orlando, Florida, USA). This work was supported by the ASRTD at University of Engineering and Technology, Taxila and Xiamen University Malaysia (XMUM).}


\conflictsofinterest{The authors declare that there is no conflict of interest regarding this publication.} 




\reftitle{References}





\end{document}